\documentclass{article}

\usepackage{amssymb}
\usepackage{psfrag} 
\usepackage{latexsym} 
\usepackage{graphicx} 

\begin{document}



\begin{center}
{\Large Effect of Kondo resonance on optical third harmonic generation}\\
\vspace{3mm}
 S.A. Jafari$^{1,2}$, T. Tohyama$^1$, S. Maekawa$^{1,3}$\\
 {\em $^1$Institute for Materials Research, Sendai 980-8577, Japan\\
 $^2$Department of Physics, Isfahan University of Technology, 
 Isfahan 84156, Iran\\
 $^3$CREST, Japan Science and Technology Agency, Kawaguchi 332-0012, Japan\\}
\end{center}

\begin{abstract}
We use the method of dynamical mean field thoery, to study the 
effect of Kondo resonance on optical third harmonic generation (THG) spectra of 
strongly correlated systems across the metal-insulator transition.
We find that THG signals are proportional to the quasiparticle 
weight $z$ of the Kondo peak, and are precursors of Mott-Hubbard gap formation. 
\end{abstract}



\section{Introduction}
One of the earliest and most important triumphs of the 
Dynamical mean field theory (DMFT) \cite{KotliarRMP} 
was to elucidate the nature of metal-to-insulator transition (MIT) driven by 
strong electron correlation. It was demonstrated that this transtion
is accompanied by a Kondo resonance at the Fermi surface \cite{xyzhang93}, 
width of which is proportional to the quasiparticle weight $z$, and
approaches to zero as the Mott insulating side ($U\to U_c^-$) is 
approached from the metallic side. 
Therefore in nonlinear optical spectra, we expect
some signals that vanish in the limit $U\to U_c^-$, as a 
result of qusiparticle weight $z\to 0$.

   Equivalently one can employ the pressure to 
tune $U_c$ and hence switch from Mott insulating side to 
a regime in which such  signals
arise due to Kondo peak. This is a unique feature of correlated insulators, 
that has no analogue in band-insulators. In case of band insulators, to
generate quasiparticle states in the mid-gap, one may need to dope
them, while in Mott insulators such mid-gap states can be 
created by applying pressure to a Mott insulator which is in the onset 
of MIT. Possible examples would be a class of materials known as 
Kondo insulators \cite{HRK-kondo}. (A typical example of this family 
is YbB$_{12}$ \cite{Susaki}).

\section{Method of calculation}
   One of the applications of nonlinear opitcal materials is the
third harmonic generation (THG) \cite{BoydBook}, that requires materials
capable of providing strong enough nonlinear optical signals. To study this process
in correlated electronic systems, we have developed 
the theory of nonlinear optical response in the context of DMFT \cite{JafariDMFT-THG}
which is also a suitable theoretical tool to study the effect of Kondo resonance in 
nonlinear spectra. In $d\to\infty$ limit where DMFT becomes exact, vertex 
corrections to current operators identically vanish \cite{PeskinBook} so that the
THG sucseptibility within DMFT approximation becomes \cite{JafariDMFT-THG}
\begin{equation} 
   \chi^{\rm THG}(\nu) \!=\!  
   \frac{\tilde t^4}{6\pi\nu^4} \!\int\!  
   \frac{d\omega d\varepsilon D(\varepsilon)}{\xi_0-\varepsilon} 
   \frac{1}{\xi_1-\varepsilon} 
   \frac{1}{\xi_2-\varepsilon} 
   \frac{1}{\xi_3-\varepsilon}, 
   \label{thgConvolution.eq}
\end{equation}
where
$\xi_j=\omega+j\nu-\Sigma_R(\omega+j\nu)+i\vert\Sigma_I(\omega+j\nu)\vert$ 
for $j=0,1,2,3$, and $D(\varepsilon)=2\sqrt{1-\varepsilon^2}/\pi$ is the semicircular
DOS corresponding to the normalized hopping $\tilde t=1/2$. 

   In the above formula, (i) the integration over $\varepsilon$  
corresponds to summation over intermediate states in conventional 
expressions \cite{BoydBook} which are usually used for 
systems with discrete energy levels, and 
(ii) the matrix element effects are encoded in $\Sigma(\omega)$, 
real and imaginary part of which have been denoted by $\Sigma_R$ 
and $\Sigma_I$, respectively. 
{\em It is very crucial to note that we have used absolute value  
of the imaginary part of the self-energy}. This is indeed 
a necessary step to transform from time-ordered four-current, 
to fully retarded one\cite{JafariJPSJ}.

We obtain self-energy $\Sigma$ by solving the iterated perturbation theory 
equations of DMFT for the semi-circular density of states \cite{ThanksRozenberg}. 
In DMFT approximation, this quantity is purely local. 

\begin{figure}[t] 
   \centering 
   \includegraphics[height=0.55\textwidth,angle=0]{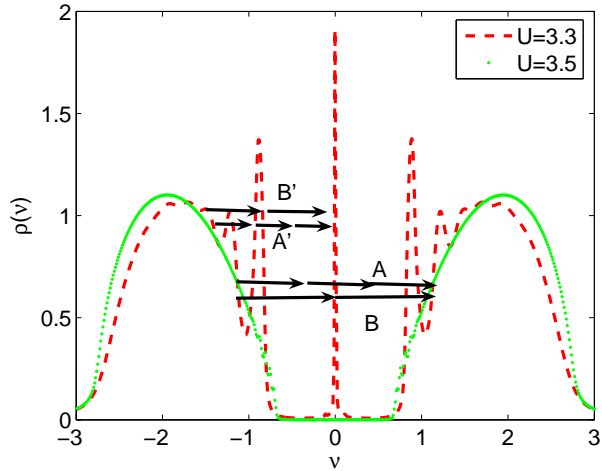} 
   \caption {Density of states within DMFT at the onset of transition to
   insulating state (dashed line) and in the Mott insulating state (dotted line).
   {\bf A} and {\bf A'} represent schematically the three-photon resonance processes
   in $\chi^{\rm THG}$ of Fig. \ref{thg-kondo.fig}. Such three-photon resonances
   appear as peaks {\bf A} and {\bf A'} in 
   Fig. \ref{thg-kondo.fig}, as explained in Ref. \cite{JafariDMFT-THG}.
   Similarly {\bf B} and {\bf B'} in this figure schematically represent two-photon 
   resonances denoted by {\bf B} and {\bf B'} in Fig. \ref{thg-kondo.fig}. }
   \label{MITdos.fig} 
\end{figure}

The critical value of Hubbard $U$ corresponding to
semicircular Bethe lattice is $U_c\sim 3.35$, above which
we are in Mott insulating phase (Fig. \ref{MITdos.fig}). As can be seen in 
Fig. \ref{MITdos.fig}, the sharp Kondo resonance state at Fermi surface
in the onset of transition to Mott insulating phase,
is characteristic of correlation driven MIT.

\section{Results}
\begin{figure}[t] 
   \centering 
   \includegraphics[height=0.55\textwidth,angle=0]{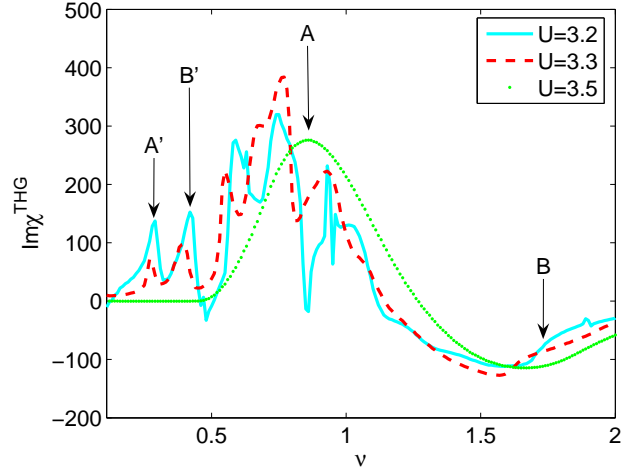} 
   \caption {Imaginary part of $\chi^{\rm THG}$ across the Mott
   metal-to-insulator transition. Critical value of $U$ is $\sim 3.35$. 
   Solid line corresponds to $U=3.2$, 
   dashed line corresponds to $U=3.3$, 
   and dotted line corresponds to $U=3.5$, which is in the Mott insulating 
   side of MIT transition.
   The onset of three-photon absorption for $U=3.5$ is where imaginary
   part of $\chi^{\rm THG}$ starts to become non-zero (around $\nu\sim 0.5$).} 
   \label{thg-kondo.fig} 
\end{figure}

   Fig. \ref{thg-kondo.fig} shows the imaginary part of the THG susceptibility
$\chi^{\rm THG}(\nu)$ of eq. (\ref{thgConvolution.eq}) as a function of incident
photon energy $\nu$ for various values of Hubbard $U$ close to MIT. Dotted line
corresponds to $U=3.5$ which is in the Mott insulating phase. The Mott-Hubbard gap
is manifested by a region in which Im$\chi^{\rm THG}$ is identically 
zero. The onset of three-photon absorption as demonstrated
earlier \cite{JafariDMFT-THG} corresponds to the frequency where Im$\chi^{\rm THG}$
takes up, and is 1/3 of the Mott-Hubbard gap ($E_g$).

   Three-photon resonances in this formulation correspond
to the peak {\bf A} in Im$\chi^{\rm THG}$ around $\nu\sim 0.8$ (dotted line),
while two-photon features appear as a dip {\bf B} around 
$\nu\sim 1.7$ \cite{JafariDMFT-THG}.
As can be seen in Fig. \ref{thg-kondo.fig}, except for 
slight shift in the location of the dip {\bf B}, the two-photon features
are not much affected by moving to lower values of $U$ in $U\lesssim U_c$ region. 
In contrast, the three-photon features is entirely distorted
as we cross the $U_c$ and enter the Kondo regime of $U\lesssim U_c$.

Below the critical value $U_c\sim 3.35$ the system is in 
metallic side, with a sharp Kondo peak present at the Fermi
surface \cite{xyzhang93}. For $U=3.3$ (dashed line) just below
the critical value, there appear peaky structures below $\nu\sim 1$. Among 
these structures, weak features denoted by {\bf A'} and {\bf B'} 
in Fig. \ref{thg-kondo.fig} start to appear in the THG 'gap' region 
($0<\nu<E_g/3$). These features seem to correspond to three-photon and two-photon 
rosonances between the lower Hubbard band and the Kondo resonance 
state (schematically depicted in Fig. \ref{MITdos.fig}), as they vanish
in $z\to 0$ limit.
Moving further away from MIT, quasiparticle weight $z$ grows and hence {\bf A'}
and {\bf B'} features also start to grow proportional to $z$. 
To see this proportionality, note 
that $\chi^{\rm THG}$ in eq. (\ref{thgConvolution.eq})
is dominated by $D(\varepsilon)$ effect, and hence in the limit $z\to 0$
one can replace the Kondo peak with a sharp peak of strength $z$. Then decomposing
to partial fractions, each with a single 
$(\xi_j-\varepsilon)$ denominator in eq. (\ref{thgConvolution.eq}),
one can see that $j-$photon
features for $j=1,2,3$ will be proportional to $z$.

   Implication of this observation for Kondo insulators is that, an external 
pressures on the order of a fraction of GPa can lead to instability in the 
functioning of optical devices built on them by filling in the gap 
and hence reducing the contrast of the output signal.

\section{Acknowledgements} 
S.A.J. was supported by JSPS fellowship P04310.
We wish to thank M.J. Rozenberg for using his code in solving
DMFT equations. This work was supported by Grant-in-Aid for 
Scientific Research from the MEXT, and NAREGI project.



\end{document}